\documentclass[preprint,12pt]{elsarticle}
\usepackage{lineno,hyperref}
\usepackage{amsmath,amssymb}
\usepackage[english]{babel}
\usepackage{bm}
\usepackage{color}
\usepackage{graphicx}
\usepackage{subfigure}
\newcommand{\angstrom}{\mbox{\normalfont\AA}}

\usepackage[normalem]{ulem}
\usepackage{color}
\definecolor{ascol}{rgb}{0.7,0, 0}

\journal{Computational Materials Science}

\begin{document}

\begin{frontmatter}

\title{Improving accuracy of interatomic potentials: more physics or more data? A case study of silica.}

\author[mainaddress]{Ivan S. Novikov}

\author[mainaddress]{Alexander V. Shapeev\corref{correspondingauthor}}
\address[mainaddress]{Skolkovo Institute of Science and Technology, Skolkovo Innovation Center, Nobel St. 3, Moscow 143026, Russia}

\cortext[correspondingauthor]{Corresponding author}
\ead{a.shapeev@skoltech.ru}

\begin{abstract}
In this paper we test two strategies to improving the accuracy of machine-learning potentials, namely adding more fitting parameters thus making use of large volumes of available quantum-mechanical data, and adding a charge-equilibration model to account for ionic nature of the SiO$_2$ bonding.
To that end, we compare Moment Tensor Potentials (MTPs) and MTPs combined with the charge-equilibration (QEq) model (MTP+QEq) fitted to a density functional theory dataset of $\alpha$-quartz SiO$_2$-based structures.
In order to make a meaningful comparison, in addition to the accuracy, we assess the uncertainty of predictions of each potential.
It is shown that adding the QEq model to MTP does not make any improvement over the MTP potential alone, while adding more parameters does improve the accuracy and uncertainty of its predictions.
\end{abstract}


\end{frontmatter}

%\linenumbers

\section{Introduction}

Oxides represent one of the most important class of functional materials deriving their unique properties from the nature of ionic bonding of oxygen. 
Silicon dioxide (SiO$_2$, silica), although not rated as a functional material, has been extensively studied as it has many industrial applications including semiconductors \cite{satterthwaite2016-Si-experiment}, metal casting \cite{rao2007-metal-casting}, and in the production of glass \cite{stachurski2015-amorphous-solids}, to name a few.
%All the facts mentioned above illustrate an importance of SiO$_2$ study under various conditions (temperatures and pressures).
In addition to experimental works \cite{levien1980-quartz-pressure, strauch1993-quartz-experiment, ghita2011-Si-surface}, there has been extensive efforts in studying SiO$_2$ computationally.
Density functional theory (DFT) is able to correctly describe ionic bonds and Coulombic interaction and hence provides an accurate description of interatomic interaction in the SiO$_2$ system.
However, DFT is too computationally expensive to model some of the critical properties of SiO$_2$ such as the solidification of molten silicon, for which the interaction model has to be several orders of magnitude more computationally efficient.

Empirical interatomic potentials has thus been the only alternative to DFT for conducting large-scale simulations of SiO$_2$.
Examples of such potentials are the Stillinger-Weber (SW) potential \cite{watanabe1999-Si-SW, ganster2010-Si-SW-strain-diffusion} and the Tersoff potential \cite{billeter2006-Si-AI-Tersoff}.
These potentials have a relatively simple functional form for the short-range interatomic interactions and they do not explicitly capture long-range Coulombic interactions in oxides.
For example, the SW potential is a pair-interaction potential with a three-body term penalizing the bond angles that differ from the ones expected to occur in SiO$_2$.
Many efforts have been made to put more physics into the model to make it more accurate.
For instance, fixed-charge pair potentials combining short-range and long-range interactions have been proposed in \cite{williams1984-Coul-fixed-q, vanbeest1990-Si-Coul-fixed-q}.
However, such the potentials cannot readjust to match the electrostatic environment.
In \cite{tangney2002-Si-opt-q, kermode2010-Si-polar} the above fixed-charge potentials were extended.
In those interatomic interaction models the charges became the parameters which were optimized and the effects of dipole polarization of the oxygen ions were taken into account.
The charge-equilibration (QEq) model is a more sophisticated model proposed by Rappe and Goddard \cite{rappe1991-QEq}.
QEq allows the charges to respond to changes in the electrostatic environment.
Some interatomic interaction models, such as modified Tersoff \cite{yasukawa1996-SiO2-Tersoff-QEq}, reactive force field (ReaxFF) \cite{vanduin2003-SiO-reaxff} and charge optimized many-body (COMB) potential \cite{yujianguo2007-Si-COMB} were constructed on the basis of the QEq model.
These potentials have been successfully used in the description of SiO$_2$.
However, application of these potentials to large-scale systems may be limited due to the fact that the QEq method requires a significantly larger computational effort than conventional empirical potentials.

Another research direction intended to increase the accuracy of the interatomic potentials is the so-called machine-learning interatomic potentials \cite{behler2007-NNP,bartok2010-GAP,szlachta2014-W-GAP,boes2016-Au-NNvsReaxFF, shapeev2016-mtp, gubaev2018-chemoinformatics, gubaev2018-mtp-multicomponent, artrith2015-Cu-Au-NNP, behler2011-NNP, behler2014-NNP, dolgirev2016-ML, gastegger2015-NNP-organic, manzhos2015-NN-PES, behler2015-water, lubbers2018-DNN, smith2017-NNP, kolb2017-DFT-ML-coupling, deringer2016-GAP-carbon, deringeri2018-GAP-boron, grisafi2018-ML-tensorial-properties, thompson2015-automated, botu2015-MLIP, devita2015-LOTF, kruglov2017-energy-free}.
Ideologically, they are different from the empirical interatomic potentials in the way that machine-learning potentials attempt to increase accuracy not by putting more physics into the model, but through a flexible functional form that allows large amounts of DFT data to be used for the fitting.
The first work on this topic was published by Behler and Parinello \cite{behler2007-NNP}.
They constructed a neural network potential (NNP) and successfully applied it to modelling of silicon; in particular, their potential predicted the radial distribution function of a silicon melt at 3000 K with high accuracy.
In \cite{bartok2010-GAP} on the basis of the idea of Gaussian process regression, the Gaussian approximation potential (GAP) was proposed and successfully applied for prediction of various properties of carbon, silicon and germanium. Thereafter, many works appeared that propose or validate interatomic potentials based on neural networks \cite{artrith2015-Cu-Au-NNP, behler2011-NNP, behler2014-NNP, boes2016-Au-NNvsReaxFF, dolgirev2016-ML, gastegger2015-NNP-organic, manzhos2015-NN-PES, behler2015-water, lubbers2018-DNN, smith2017-NNP, kolb2017-DFT-ML-coupling,schutt2017schnet},
Gaussian processes \cite{szlachta2014-W-GAP, deringer2016-GAP-carbon, deringeri2018-GAP-boron, grisafi2018-ML-tensorial-properties} and other methods \cite{shapeev2016-mtp, gubaev2018-chemoinformatics, gubaev2018-mtp-multicomponent, thompson2015-automated}. 
In the above works only short-range interaction was taken into account.
Behler and his collaborators extended their NNP by including electrostatic interactions explicitly in the functional form \cite{artrith2011-NNP-charges} and testing it for ZnO, not reporting, however, that it improves the accuracy of their potential.
A similar model directly predicting atomic charges has recently been proposed in \cite{nebgen2018-charge-assignment}.
The field of machine-learning interatomic potentials is ideologically and methodologically close to the field of machine-learning cheminformatics---developing of models predicting the properties of molecules and materials directly, without a molecular simulation \cite{rupp2012-coulomb-matrix,huo2017-mbtr,snyder2012-ML-for-DFT,de2016-soap-cheminformatics,faber2017-cheminformatics,hansen2015-cheminformatics,huang2016-cheminformatics,gilmer2017-ceminformatics,schutt2017-DTNN}.

Many studies on machine-learning potentials report \cite{behler2007-NNP, bartok2010-GAP, szlachta2014-W-GAP, dolgirev2016-ML, smith2017-NNP, deringer2016-GAP-carbon} that these potentials are more accurate than off-the-shelf empirical potentials.
A conceptually interesting study was \cite{szlachta2014-W-GAP}, where the authors considered a growing set of quantities of interest (phonons, elastic constants, defects, surfaces, etc.) and show that it is possible to construct a series of potentials that reproduce this growing set without losing accuracy by increasing the number of parameters in a potential.
Another interesting paper is \cite{boes2016-Au-NNvsReaxFF}, where the authors show that when the potentials are fitted on a very large configurational space (containing bulk, surfaces, clusters, etc.) for pure gold, a machine-learning potential was significantly more accurate than ReaxFF.
Gold has largely delocalized metallic bonds that are hard to represent explicitly in an empirical potential with high accuracy, however, SiO$_2$ has ionic/covalent bonding which empirical potentials are expected to represent sufficiently accurately.
It would hence be interesting to repeat a similar study for SiO$_2$, however, this goes beyond the scope of the present study because the parametrization of empirical potentials is methodologically very different from the parametrization of machine-learning potentials.

The main purpose of this work was to assess and compare two possible strategies for improving the accuracy of a machine-learning potential.
The first strategy is adding more parameters to a machine-learning potential thereby utilizing large amounts of available DFT data.
The second approach is to add a charge-equilibration model, and hence capture the ionic nature of SiO$_2$ bonding better.
To that end, we compare two classes of models, the Moment Tensor Potential (MTP) and the MTP with the additional charge-equilibration model (MTP+QEq).
MTP was first proposed in \cite{shapeev2016-mtp} for the case of single-component materials and extended in \cite{gubaev2018-chemoinformatics,gubaev2018-mtp-multicomponent} to multi-component materials.

This paper is organized as follows. In Section \ref{sec:meth} we describe the methodology namely, the interatomic potentials, their fitting, and our uncertainty quantification method.
In Section \ref{NumExp} we present and discuss the results of numerical experiments.
In particular, we compare the training errors in energies, forces and stresses for different models, as well as elastic constants, vacancy formation energies (VFEs), phonon spectra and radial distribution functions (RDFs) obtained by these models. Finally, in Section \ref{Discussion} we give the concluding remarks. 

\section{Methodology}\label{sec:meth}

In order to address the main question of this study, namely to what extent the accuracy of the description of ionic bonds can be improved by adding a charge-equilibration (QEq) model, we introduce the MTP and the combined MTP+QEq model (Section \ref{sec:meth:IIMs}).
Since our main focus is the machine-learning potentials, we consider a local optimization method for finding the parameters (Section \ref{sec:meth:fitting}).
Finally, in order to better analyze the results, we introduce an uncertainty quantification method in Section \ref{sec:meth:uncertainty}.

\subsection{Interatomic potentials} \label{sec:meth:IIMs}

%We start by formulating an SW potential in the form similar to the one from \cite{ganster2010-Si-SW-strain-diffusion}.
%The potential has originally been developed to describe bulk silicon \cite{stillinger1985-SW-origin}.
%It expresses local interaction of atoms and does not explicitly account for the Coulombic long-range interactions. 
Let $\bm x=\{(x_i,z_i) : i=1,\ldots,n\}$ be a configuration with $n$ atoms, each atom is encoded by its position $x_i$ and atomic type $z_i$. We assume that the $i$-th atom interacts with its neighbors and we refer to the $j$-th atom as the \emph{neighbor} of the $i$-th atom if the distance between them is not greater than a cutoff radius $R_{\rm  cut}$. The locality of interaction is expressed by expanding the total interaction energy as a sum of contributions of individual atoms: $E(\bm x) = \sum_i V_i := \sum_i V(\bm r_i)$, where ${\bm r_i} = (r_{i\;\!1}, \ldots, r_{ij}, \ldots, r_{i\;\!n})$ is the neighborhood of the $i$-th atom, $r_{ij} = x_j - x_i$ is the position of the $j$-th atom relative to the $i$-th atom.

\begin{figure}[h!] \begin{center}
\includegraphics[width=4.0in, height=2.5in, keepaspectratio=false]{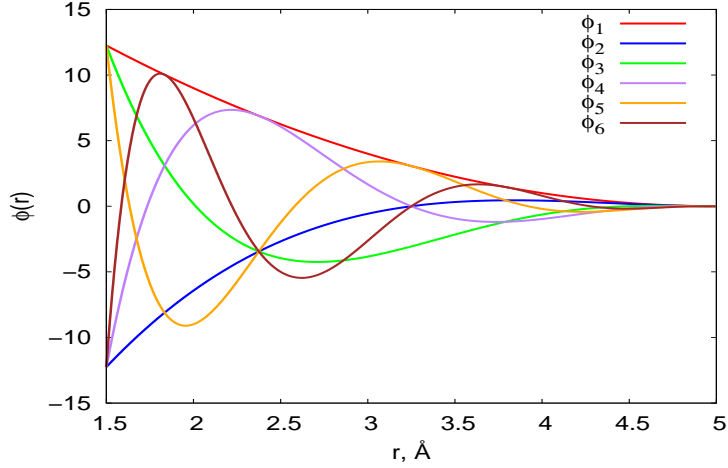}
\begin{center}
\caption{\label{fig:Chebyshev} Radial basis $\varphi_1,\ldots,\varphi_6$.}
\end{center}
\end{center} \end{figure}

Now we introduce MTP, first proposed in \cite{shapeev2016-mtp} and then generalized to multiple components in \cite{gubaev2018-mtp-multicomponent,gubaev2018-chemoinformatics}. It has the following form:
\begin{align} \label{MTPrepresentation}
\displaystyle 
V^{\rm MTP}_i &:= \sum _{\alpha} \xi_{\alpha} B_{\alpha} (\bm r_i),
\end{align}
where $\xi_{\alpha}$ are the parameters to be fitted and $B_{\alpha} (\bm r_i)$ are the basis functions. 
%These functions depend on the atomic types, on the distances between atoms and on the parameters $c_{\mu, z_i, z_j}$ that should be defined during fitting the potential. In the above notations $\mu$ is one of the numbers that define moment tensor degrees (see \cite{Gubaev} for details). 
In order to define these functions we introduce the so-called \emph{moment tensor} descriptors: 
\begin{equation}\label{Moments}
M_{\mu,\nu}(\bm r_i)=\sum_{j} f_{\mu}(|r_{ij}|,z_i,z_j) \underbrace {r_{ij}\otimes...\otimes r_{ij}}_\text{$\nu$ times},
\end{equation}
where the symbol ``$\otimes$'' stands for the outer product of vectors and therefore the angular part $r_{ij}\otimes...\otimes r_{ij}$ resembles the moments of inertia, $f_{\mu}(|r_{ij}|,z_i,z_j)$ is the radial part of the following form:

$$f_{\mu}(|r_{ij}|,z_i,z_j) = \sum_{\beta} c^{(\beta)}_{\mu, z_i, z_j} \varphi_\beta (|r_{ij}|),$$

\noindent in which $c^{(\beta)}_{\mu, z_i, z_j}$ are the parameters to be fitted and $\varphi_\beta (|r_{ij}|)$ are the radial basis functions \eqref{RadialBasis}:

\begin{align} \label{RadialBasis}
\displaystyle
\varphi_{\alpha}(r) =\begin{cases}
T^{\alpha} (r) (R_{\rm cut} - r)^2 & r<R_{\rm cut} \\
0 & r \geq R_{\rm cut},
\end{cases}
\end{align}
where $T^{\alpha} (r)$ is the Chebyshev polynomial of degree $\alpha$ on the interval $[R_{\rm min},R_{\rm cut}]$, the term $(R_{\rm  cut}-r)^2$ is introduced to ensure a smooth cutoff to 0 at $r\geq R_{\rm cut}$, and $R_{\rm min}$ is some lower bound on minimal interatomic distances.
For illustration, we plot the first six radial basis functions on the interval $[1.5\,\angstrom, 5\,\angstrom]$ in Figure \ref{fig:Chebyshev}.
We construct our basis functions $B_{\alpha}$ as all possible contractions of the \emph{moment tensor} descriptors \eqref{Moments} to scalar, e.g.:
\begin{align*}
B_0(\bm r_i) &= M_{0,0}(\bm r_i),
\\
B_1(\bm r_i) &= (M_{1,2}(\bm r_i) M_{0,1}(\bm r_i)) \cdot M_{1,1}(\bm r_i),
\\
B_2(\bm r_i) &=M_{0,2}(\bm r_i) : M_{2,2}(\bm r_i),
\\
&...
\end{align*}
where ``$\cdot$'' is the dot product, ``:'' is the Frobenius product, $M_{0,0}(\bm r_i)$ is a scalar ($\nu = 0$), $M_{0,1}(\bm r_i)$ and $M_{1,1}(\bm r_i)$ are vectors ($\nu = 1$), while $M_{0,2}(\bm r_i)$, $M_{1,2}(\bm r_i)$ and $M_{2,2}(\bm r_i)$ are matrices ($\nu = 2$).
We denote the free parameters of MTP by ${\bm {\theta}}^{\rm MTP} := (\xi_{\alpha}, c^{(\beta)}_{\mu, z_i, z_j})$. The total interaction energy is $E^{\rm MTP} = E({\bm {\theta}}^{\rm MTP}; \bm x) = \sum_{i=1}^{n} V^{\rm MTP}_i$.

Next we describe the QEq model. This model was proposed in \cite{rappe1991-QEq} and we use it in the following form:
\begin{align} \label{QEQrepresentation}
\displaystyle
E^{\rm QEq} &:= E (\bm {\theta}^{\rm QEq}; \bm x, \bm q) = \sum _{i=1}^n 
\Bigl( \chi_{z_i} q_i + \dfrac{J_{z_i} q_i^2}{2} + \sum _{j<i} \dfrac{q_i q_j}{|r_{ij}|} \Bigr),
\end{align}

\noindent where $\chi_{z_i}$ is the electronegativity of the atom of type $z_i$, $J_{z_i}$ describes repulsion between two electrons, $q_i$ and $q_j$ are the partial charges of the $i$-th and the $j$-th atom, respectively. The third sum in \eqref{QEQrepresentation} describing the Coulombic interaction converges only conditionally, therefore we use the Ewald summation \cite{ewald1921} for its calculation. We denote the free parameters of the QEq model by ${\bm {\theta}}^{\rm QEq} := (\chi_{z_i}, J_{z_i})$ and the collection of partial charges in a configuration by $\bm q := (q_1,\ldots,q_n)$.
It should be emphasized that \eqref{QEQrepresentation} is a long-range interaction model with no cutoff radius.

We find the partial charges $q_i$ by solving the following optimization problem for each configuration $\bm x$ occurring in a simulation:
\begin{equation} \label{ChargeEquilibration}
\begin{array}{c}
\displaystyle \bm q^* = \underset{\bm q}{\operatorname{arg\,min}}\, E (\bm {\theta}^{\rm QEq}; \bm x, \bm q)
\\
\displaystyle \mbox{subject to} ~\sum _{i=1}^n q_i = 0.
\end{array}
\end{equation}
This expresses equilibration of charges, see \cite{rappe1991-QEq} for details.

The second interatomic interaction model used in this work is, thus, the combination of the MTP and the QEq model: 
\begin{equation} \label{MTP+QEq}
\displaystyle
E^{\rm MTP+QEq} := E(\bm {\theta}^{\rm MTP+QEq}; \bm x, \bm q) = E({\bm {\theta}}^{\rm MTP}; \bm x) + E(\bm {\theta}^{\rm QEq}; \bm x, \bm q),
\end{equation}

\noindent where $\bm {\theta}^{\rm MTP+QEq} := ({\bm {\theta}}^{\rm MTP}, {\bm {\theta}}^{\rm QEq})$. 

\subsection{Fitting}\label{sec:meth:fitting}

We now describe the optimization problems for finding the free parameters of the models described above.
Let us consider a training set which consists of $K$ configurations $\bm x^{(k)}$ ($k=1, \ldots, K$).
Suppose also that we have the DFT energies
$E^{{\rm DFT},(k)}$, forces $\bm f^{{\rm DFT},(k)}$, and stresses $\bm \sigma^{{\rm DFT},(k)}$, where
by $\bm f^{{\rm DFT},(k)}$ we mean the collection of forces on each atom of the $k$-th configuration and by $\bm \sigma^{{\rm DFT},(k)}$ the collection of stresses of the $k$-th configuration.
In order to find the free parameters of the MTP we solve the following optimization problem:
\begin{equation} \label{LocalFitting}
\begin{array}{r@{}l}
\displaystyle
L({\bm {\theta}}^{\rm MTP}) = \sum _{k=1}^K \Bigl[ &w_{\rm e} \left(E^{{\rm DFT},(k)} - E({\bm {\theta}}^{\rm MTP}; \bm x^{(k)}) \right)^2 \\
\displaystyle
&+ w_{\rm f} \sum _{i} \left| f^{{\rm DFT}, (k)}_i - f_{i}({\bm {\theta}}^{\rm MTP}; \bm x^{(k)}) \right|^2 \\
\displaystyle
&+ w_{\rm s} \sum _{i,j=1}^{3} \left( \sigma^{{\rm DFT},(k)}_{ij} - \sigma_{ij}({\bm {\theta}}^{\rm MTP}; \bm x^{(k)}) \right)^2 \Bigr]
\to \operatorname{min},
\end{array}
\end{equation}
where $w_{\rm e}$, $w_{\rm f}$ and $w_{\rm s}$ are some nonnegative weights.
We solve a similar problem in order to find the MTP parameters. 

Because of partial charges in the combined model \eqref{MTP+QEq} we apply a slightly different algorithm for finding the free parameters of MTP+QEq.
Before each iteration of minimizing the objective function
\begin{equation} \label{NonlocalFitting}
\begin{array}{r@{}l}
\displaystyle
L({\bm {\theta}}^{\rm MTP+QEq}) = \sum _{k=1}^K \Bigl[& w_{\rm e} \left(E^{{\rm DFT},(k)} - E({\bm {\theta}}^{\rm MTP+QEq}; \bm x^{(k)}, \bm q^{*,(k)}) \right)^2
\\
\displaystyle 
&+ w_{\rm f} \sum _{i=1}^{N^{(k)}} \left| f^{{\rm DFT}, (k)}_i - f_{i}({\bm {\theta}}^{\rm MTP+QEq}; \bm x^{(k)}, \bm q^{*,(k)}) \right|^2
\\
\displaystyle 
&+  
w_{\rm s} \sum _{i,j=1}^{3} \left( \sigma^{{\rm DFT},(k)}_{ij} - \sigma_{ij}({\bm {\theta}}^{\rm MTP+QEq}; \bm x^{(k)}, \bm q^{*,(k)}) 
\right)^2 \Bigr]
\to \operatorname{min},
\end{array}
\end{equation}
we optimize the charges $q^{*,(k)}$ for each configuration by solving the problem \eqref{ChargeEquilibration}. 
Thus, we have different equilibrated charges $q^{*,(k)}$ during each iteration of solving \eqref{NonlocalFitting}.

Since the models depend nonlinearly on the parameters $\bm \theta$, we use a quasi-Newton optimization method, namely, the Broyden--Fletcher--Goldfarb--Shanno algorithm (BFGS).
To that end, we explicitly implemented the gradients of the loss function with respect to the parameters $\bm \theta$ for each of the methods.
In particular, for MTP, we have implemented an efficient back-propagation algorithm which has a favorable scaling when the number of parameters is large.

\subsection{Uncertainty quantification}\label{sec:meth:uncertainty}

The fitted potential is a random quantity: it depends on the random training set and/or a particular local minimum that the optimization routine has found.
Therefore the predictions of such potentials are also, strictly speaking, random.
In order to quantify such randomness (uncertainty), we fit an ensemble of potentials of each type starting from random initial values of parameters and analyze the distribution of predictions by each type of potential, not just a single value of the ``best'' potential.
In particular, we analyze the standard deviation (sometimes called \emph{predictive variance}) of predictions of the ensemble of potentials and compare it to the actual error.
As we will see, in our tests the standard deviation gives a good estimation of the actual error in all the quantities of interest considered in this study.
Such technique of estimating the uncertainty of predictions is known as \emph{query by committee} \cite{settles2012-AL}.

\section{Numerical Testing} \label{NumExp}

We fit two potentials, MTP and MTP+QEq, and test how well they predict the elastic constants, VFE, phonon spectrum, and RDFs of the $\alpha$-quartz SiO$_2$. $\alpha$-quartz is the stable crystalline structure of SiO$_2$ at normal temperature and pressure.

\subsection{Training dataset}

We composed the training dataset by perturbing and introducing defects to the $\alpha$-quartz SiO$_2$. This structure has a trigonal unit cell with the lattice parameters $a = b = 5.022 ~\angstrom$, $c = 5.551 ~\angstrom$, $\alpha = \beta = 90^{\circ}$, and $\gamma = 120^{\circ}$. The unit cell contains 3 atoms of silicon and 6 atoms of oxygen.

We first generated 22 (all possible) supercells with 9, 18, 27 and 36 atoms by replicating the unit cell in different axes and applying shear. Next, in addition to pristine crystals, we generated a number of structures with a single O-vacancy defects, which were then relaxed (equilibrated). We used the VASP code \cite{kresse1993-VASP,kresse1994-VASP,kresse1996A-VASP,kresse1996B-VASP} for DFT calculations with the PBE functional \cite{perdew1996-DFT-PBE}, the PAW pseudopotentials \cite{blochl1994-DFT-PAW}, k-point meshes were equivalent to the $4 \times 4 \times 3$ k-point mesh in the unit cell and a cutoff energy was 400 eV. After the configurations were relaxed, we randomly displaced every atom in each configuration by about $0.1 ~\angstrom$. From each ``undisplaced'' configuration several configurations with displaced atoms were generated.
%We had to do it because one of the physical characteristics for comparison is phonon spectrum calculated by different interatomic interaction models. 
In addition, in order to predict elastic constants, we added 13 more configurations to the training dataset: 12 of them are relaxed configurations without atomic defects and with lattice vectors extended, compressed, or sheared by 2\%, i.e., two configurations with extensions/compressions along each of six directions ($xx$, $yy$, $zz$, $yz$, $xz$, $xy$) and one of them is the relaxed configuration without any defects and extensions/compressions of lattice vectors.
Thus, our dataset contains 418 SiO$_2$ atomic configurations with oxygen vacancies, random displacements of atomic positions and shear/compression.

%For each configuration we calculated energies, forces and stresses using VASP. In order to parametrize our potentials, i.e., to solve the problems \eqref{LocalFitting}, \eqref{NonlocalFitting} we use the algorithm for the objective functions local minima search, namely, Broyden--Fletcher--Goldfarb--Shanno algorithm (BFGS). This algorithm belongs to the class of quasi-newton methods. We start from random initial guess for fitting of each local potential. 

\subsection{Comparison of Potentials}

We fit two types of potentials: MTP \eqref{MTPrepresentation} and MTP+QEq \eqref{MTP+QEq}.
For all the potentials we choose $R_{\rm min} = 1.4 \,\angstrom$, $R_{\rm cut} = 5 \,\angstrom$, and eight radial functions \eqref{RadialBasis}. 
We consider the MTPs with 9, 29 and 92 basis functions $B_{\alpha}$. We denote these potentials by MTP$_1$, MTP$_2$ and MTP$_3$, respectively.
The total number of free parameters in these potentials are 150, 250, and 500, respectively.
MTPs in the MTP+QEq models contain 9 and 29 basis functions, thus we consider MTP$_1$+QEq and MTP$_2$+QEq models. The weights in the objective functions \eqref{LocalFitting}, \eqref{NonlocalFitting} were $w_{\rm e} = 1$, $w_{\rm f} = 10^{-2} \,\angstrom^2$, and $w_{\rm s} = 10^{-3}$.

%\begin{table}
%	\begin{center}
%		\begin{tabular}{|c|c|c|c|c|} \hline 
%			Potential & energy error & force error & stress error & Si partial charge \\
%			& meV/atom & meV/\angstrom\ (\%) & GPa ($\%$) & in ideal crystal \\ \hline
%			SW & 2.61 $\pm$ 0.44 & 173.9 $\pm$ 14.2 & 0.42 $\pm$ 0.09 & - \\ 
%			& & 9.4 $\pm$ 0.8 \% & 16.4 $\pm$ 3.4 \% & \\ \hline
%			SW+QEq & 2.47 $\pm$ 0.33 & 170.0 $\pm$ 13.8 & 0.41 $\pm$ 0.08 & 0.74 $\pm$ 0.25 \\ 
%			& & 9.3 $\pm$ 0.7\% & 16.3 $\pm$ 3.3 \% & \\ \hline
%			MTP & 1.63 $\pm$ 0.09 & 90.5 $\pm$ 4.7 & 0.20 $\pm$ 0.02 & - \\ 
%			& & 4.9 $\pm$ 0.3 \%~\, & 8.7 $\pm$ 0.9 \% & \\ \hline
%		\end{tabular}
%		\caption{\label{tabl:AveTrainingErrors} Comparison of the average absolute and relative training root-mean-square errors and their standard deviations for the three models on the same dataset. In the last column the average silicon partial charge in the ideal $\alpha$-quartz crystal and its standard deviation are shown.}
%	\end{center}
%\end{table}

\begin{table}
	\begin{center}
		\begin{tabular}{|c|c|c|c|c|} \hline 
			\footnotesize Potential & \footnotesize energy error & \footnotesize force error & \footnotesize stress error & \footnotesize Si partial charge \\
			& \footnotesize meV/atom & \footnotesize meV/\angstrom\ (\%) & \footnotesize GPa ($\%$) & \footnotesize in ideal crystal \\ \hline
			\footnotesize MTP$_1$ & \footnotesize 2.66 $\pm$ \footnotesize 0.14 & \footnotesize 202.3 $\pm$ \footnotesize 7.8 & \footnotesize 0.32 $\pm$ \footnotesize 0.04 & \footnotesize - \\ 
			& & (\footnotesize 11.0 $\pm$ \footnotesize 0.4\%) & (\footnotesize 13.6 $\pm$ \footnotesize 1.7\%) & \\ \hline
			\footnotesize MTP$_1$+QEq & \footnotesize 2.71 $\pm$ \footnotesize 0.42 & \footnotesize 206.2 $\pm$ \footnotesize 20.7 & \footnotesize 0.29 $\pm$ \footnotesize 0.01 & \footnotesize 0.57 $\pm$ \footnotesize 0.11 \\ 
			& & (\footnotesize 11.1 $\pm$ \footnotesize 1.1\%) & (\footnotesize 12.3 $\pm$ \footnotesize 0.5\%) & \\ \hline
			\footnotesize MTP$_2$ & \footnotesize 1.87 $\pm$ \footnotesize 0.11 & \footnotesize 141.5 $\pm$ \footnotesize 7.0 & \footnotesize 0.20 $\pm$ \footnotesize 0.02 & - \\ 
			& & (\footnotesize 7.7 $\pm$ \footnotesize 0.4\%)~\, & (\footnotesize 8.3 $\pm$ \footnotesize 0.9\%) & \\ \hline
			\footnotesize MTP$_2$+QEq & \footnotesize 1.96 $\pm$ \footnotesize 0.18 & \footnotesize 146.0 $\pm$ \footnotesize 11.8 & \footnotesize 0.21 $\pm$ \footnotesize 0.03 & \footnotesize 0.66 $\pm$ \footnotesize 0.11 \\ 
			& & (\footnotesize 8.0 $\pm$ \footnotesize 0.6\%) & (\footnotesize 8.9 $\pm$ \footnotesize 1.1\%) & \\ \hline
			\footnotesize MTP$_3$ & \footnotesize 1.70 $\pm$ \footnotesize 0.08 & \footnotesize 130.5 $\pm$ \footnotesize 5.6 & \footnotesize 0.17 $\pm$ \footnotesize 0.01 & - \\ 
			& & (\footnotesize 7.1 $\pm$ \footnotesize 0.3\%)~\, & (\footnotesize 7.3 $\pm$ \footnotesize 0.6\%) & \\ \hline
		\end{tabular}
		\caption{\label{tabl:AveTrainingErrors} Comparison of the average absolute and relative training root-mean-square errors and their standard deviations for the five ensembles of models on the same dataset. In the last column the average silicon partial charge in the ideal $\alpha$-quartz crystal and its standard deviation are shown. MTP and MTP+QEq models with the same number of basis functions have similar accuracy. Increase in the number of basis functions in MTP improves the accuracy, whereas adding QEq to MTP does not.}
	\end{center}
\end{table}

For each model type we fit an ensemble of five potentials in order to be able to estimate uncertainty due to randomness of the fitting.
We have manually verified that in each ensemble all five potentials converged to a different minimum.
The average fitting errors and the standard deviations for each family of potentials are reported in Table \ref{tabl:AveTrainingErrors}.
We trained five ensembles of potentials: MTP$_1$, MTP$_2$, MTP$_3$, MTP$_1$+QEq and MTP$_2$+QEq. Each ensemble includes five potentials. We can see that MTP and MTP+QEq with the same number of basis functions have rather close training errors and, thus, adding QEq to MTP does not improve the accuracy. On the other hand, the increase in the number of functions $B_{\alpha}$ in MTP improves the accuracy and reduces the deviation in errors within an ensemble of potentials.
In other words, judging by training errors alone, adding more basis functions to the potential improves the accuracy, but adding the QEq model does not.
%Nevertheless, the training errors of the MTP$_3$ ensemble in comparison with the ones of the MTP$_2$ ensemble decreased not so significantly as the training errors of the MTP$_2$ ensemble in comparison with the ones of the MTP$_1$ ensemble. Due to this reason we do not consider MTP$_3$+QEq model in this paper.
%We trained 5 MTPs, 20 SW potentials and 20 SW+QEq combined models. We can see that MTP has the lowest error and also lowest uncertainty, i.e., lowest deviation between different MTP instances. Interestingly, the uncertainty in the error is very small---within a few percent of the actual error. For the family of SW potentials and SW+QEq models, in contrast, the fitting errors are about two times larger than those for the family of MTPs and the uncertainty in the error is three to four times larger than the same quantity for the MTPs.
%Another interesting observation is that the fitting errors of SW and SW+QEq are rather close to each other, in other words, adding QEq to SW does not significantly improve the error.
%For the family of SW+QEq potentials we also present the average partial charge on a Si atom in the ideal crystal.
%We observe that the fitting procedure produces physically meaningful results, although we did not extract and fit to DFT charges; however, the uncertainty in predicting the charge from total energy and forces is rather large.

In order to check the predictive power of the potentials we compared average elastic constants, VFEs, phonon spectra and RDFs calculated by these potentials to the results computed with DFT.

\begin{table}[h!]
	\begin{center}
		\begin{tabular}{|c|r|r@{\ }c@{\ }l|r@{\ }c@{\ }l|r@{\ }c@{\ }l|r@{\ }c@{\ }l|r@{\ }c@{\ }l|} \hline 
			 & \footnotesize DFT  & \multicolumn{3}{c|}{\footnotesize MTP$_1$} & \multicolumn{3}{c|}{\footnotesize MTP$_1$+QEq} & \multicolumn{3}{c|}{\footnotesize MTP$_2$}  & \multicolumn{3}{c|}{\footnotesize MTP$_2$+QEq} & \multicolumn{3}{c|}{\footnotesize MTP$_3$} \\ \hline
			\footnotesize C$_{11}$ & \footnotesize \footnotesize 91.1 & \footnotesize 82.2 &$\pm$& \footnotesize 21.5 & \footnotesize 104.1 &$\pm$& \footnotesize 4.4 & \footnotesize 96.7 &$\pm$& \footnotesize 16.9 & \footnotesize 94.8 &$\pm$& \footnotesize 11.9 & \footnotesize 88.5 &$\pm$& \footnotesize 9.2 \\ 
			\footnotesize C$_{12}$ & \footnotesize 5.9  & \footnotesize 1.5 &$\pm$& \footnotesize 15.8 & \footnotesize 23.7 &$\pm$& \footnotesize 5.2 & \footnotesize 12.1 &$\pm$&  \footnotesize 13.5  & \footnotesize 1.2 &$\pm$& \footnotesize 11.3 & \footnotesize 1.2 &$\pm$& \footnotesize 13.1 \\ 
			\footnotesize C$_{13}$ & \footnotesize 16.0 & \footnotesize 14.0 &$\pm$& \footnotesize 13.2 & \footnotesize 33.9 &$\pm$& \footnotesize 7.2 & \footnotesize 20.2 &$\pm$&  \footnotesize 10.1 & \footnotesize 5.2 &$\pm$& \footnotesize 15.2 & \footnotesize 8.4 &$\pm$& \footnotesize 10.0 \\ 
			\footnotesize C$_{14}$ & \footnotesize 15.8 & \footnotesize 13.7 &$\pm$& \footnotesize 4.1  & \footnotesize \footnotesize 10.0 &$\pm$& \footnotesize 1.6  & \footnotesize 13.7 &$\pm$& \footnotesize  4.2  & \footnotesize 14.4 &$\pm$& \footnotesize 3.7 & \footnotesize 18.7 &$\pm$& \footnotesize 4.5 \\
			\footnotesize C$_{33}$ & \footnotesize 93.8 & \footnotesize 89.2 &$\pm$& \footnotesize 13.7  & \footnotesize 108.7 &$\pm$& \footnotesize 11.9 & \footnotesize 97.7 &$\pm$& \footnotesize 14.4 & \footnotesize 83.4 &$\pm$& \footnotesize 22.1 & \footnotesize 87.8 &$\pm$& \footnotesize 8.9 \\
			\footnotesize C$_{44}$ & \footnotesize 53.2 & \footnotesize 59.6 &$\pm$& \footnotesize 7.0 & \footnotesize 63.9 &$\pm$& \footnotesize 4.1 & \footnotesize 61.2 &$\pm$&  \footnotesize 5.9 & \footnotesize 62.5 &$\pm$& \footnotesize 2.6 & \footnotesize 54.5 &$\pm$& \footnotesize 6.2 \\ 
			\footnotesize C$_{66}$ & \footnotesize 42.6 & \footnotesize 40.3 &$\pm$& \footnotesize 3.0  & \footnotesize 39.6 &$\pm$& \footnotesize 1.8 & \footnotesize 42.3 &$\pm$&  \footnotesize 5.0 & \footnotesize 46.5 &$\pm$& \footnotesize 1.2 & \footnotesize 43.6 &$\pm$& \footnotesize 3.7 \\ \hline
			\footnotesize bias & & \footnotesize 5.0 &&& \footnotesize 13.0 &&& \footnotesize 4.9 &&& \footnotesize 7.2 &&& \footnotesize 4.4 &&\\
			\footnotesize UE &&&& \footnotesize 12.8 &&& \footnotesize 6.1 &&& \footnotesize 11.0 &&& \footnotesize 12.0 &&& \footnotesize 8.5 \\
			\hline
		\end{tabular}
		\caption{\label{tabl:AveElasticConstants} Average elastic constants of SiO$_2$ $\alpha$-quartz (in GPa) calculated by DFT, MTPs and MTP+QEq models, and their standard deviations due to randomness in fitting.
			Bias is the root-mean-square deviation of the average predicted elastic constants from the reference DFT results. The last line shows the root-mean-square standard deviations (uncertainty estimate, UE) of the predictions. The errors in the average elastic constants prediction reduce while the number of basis functions in MTP increases. Adding QEq to MTP worsens the accuracy of elastic constants prediction, but the uncertainty in quantification is either close (MTP$_2$+QEq) or better (MTP$_1$+QEq) to the ones calculated on the plain MTPs.  
		} 
	\end{center}
\end{table}

The average elastic constants and their standard deviations for three ensembles of MTPs and two ensembles of MTP+QEq models are given in Table \ref{tabl:AveElasticConstants} together with their root-mean-square error compared to the reference DFT values (bias) and root-mean-square standard deviation (uncertainty due to randomness in the fitting).
One can see that the errors in the average elastic constants reduce as the number of basis functions in MTP increases.
The combination of MTP and QEq yields even worse elastic constants as compared to the plain MTP with the same number of basis functions.

\begin{table}[h!]
	\begin{center}
		\begin{tabular}{|c|r|r@{\ }c@{\ }l|r@{\ }c@{\ }l|r@{\ }c@{\ }l|r@{\ }c@{\ }l|r@{\ }c@{\ }l|} \hline 
			 & \footnotesize DFT  & \multicolumn{3}{c|}{\footnotesize MTP$_1$} & \multicolumn{3}{c|}{\footnotesize MTP$_1$+QEq} & \multicolumn{3}{c|}{\footnotesize MTP$_2$} & \multicolumn{3}{c|}{\footnotesize MTP$_2$+QEq} & \multicolumn{3}{c|}{\footnotesize MTP$_3$}\\ \hline
			\footnotesize VFE & \footnotesize 2.23 & \footnotesize 1.85 &$\pm$& \footnotesize 0.25 & \footnotesize 1.69 &$\pm$& \footnotesize 0.23 & \footnotesize 2.00 &$\pm$& \footnotesize 0.12 & \footnotesize 1.96 &$\pm$& \footnotesize 0.19 & \footnotesize 2.13 &$\pm$& \footnotesize 0.02 \\ \hline
			\footnotesize Bias &  & \footnotesize 0.38 &&& \footnotesize 0.54 &&& \footnotesize 0.23 &&& \footnotesize 0.27 &&& \footnotesize 0.10 &&\\ \hline
		\end{tabular}
		\caption{\label{tabl:VFE} Average vacancy formation energies (eV) of SiO$_2$ $\alpha$-quartz as calculated by the ensembles of MTPs and MTP+QEq models, their standard deviations. In the last line the absolute errors (eV) in vacancy formation energy calculations (biases) are presented. MTP$_1$ and MTP$_1$+QEq as well as MTP$_2$ and MTP$_2$+QEq gave close average VFEs, the uncertainty in predictions of the VFEs reduces and the accuracy of the VFEs calculation improves while increasing the number of basis functions.
		} 
	\end{center}
\end{table}

In order to obtain the average VFEs and their standard deviations we relaxed and calculated the energies of two configurations: the first configuration is the $2 \times 2 \times 2$ supercell of 72 atoms and the second configuration is the same supercell with the oxygen atom vacancy. The average VFEs and their standard deviations calculated by the five ensembles of potentials and the reference DFT VFE are presented in Table \ref{tabl:VFE}.
The findings are similar to those for elastic constants: MTP$_1$ and MTP$_1$+QEq as well as MTP$_2$ and MTP$_2$+QEq gave close average VFEs, the uncertainty in predictions of the VFEs reduces and the accuracy of the VFEs calculation improves with increasing the number of $B_{\alpha}$.

%For comparison of phonon spectra and RDFs, we choose the ``best'' MTP potential, SW potential and SW+QEq model---namely, we choose those that reproduce the elastic constants and VFE better than the other potentials of the same type. The partial charges calculated by the ``best'' QEq model for most of silicon atoms were approximately equal to 0.9 and the same values for most of oxygen atoms were approximately equal to $-0.45$.
%It is worth noting that for the structures with oxygen atom vacancies the partial charges distribution is non-trivial: the partial charges for the atoms around the vacancy are smaller than those for the atoms located farther from the vacancy. We illustrate this fact in Figure \ref{fig:Charges} for one of the structures with vacancies from the training set.   
%\begin{figure} \begin{center}
%\includegraphics[width=5.8in, height=2.6in, keepaspectratio=false]{charges_defects_new.eps}
%\caption{\label{fig:Charges} The partial charges of $\alpha$-quartz SiO$_2$ $1 \times 1 \times 4$ supercell (blue circles are the silicon atoms, red circles are the oxygen atoms) with the oxygen atom vacancy (black circle). The partial charges around the vacancy are smaller than the other partial charges.}
%\end{center} \end{figure}

\begin{figure} \begin{center}
\subfigure[]{\includegraphics[width=1.7in, height=1.7in, keepaspectratio=false]{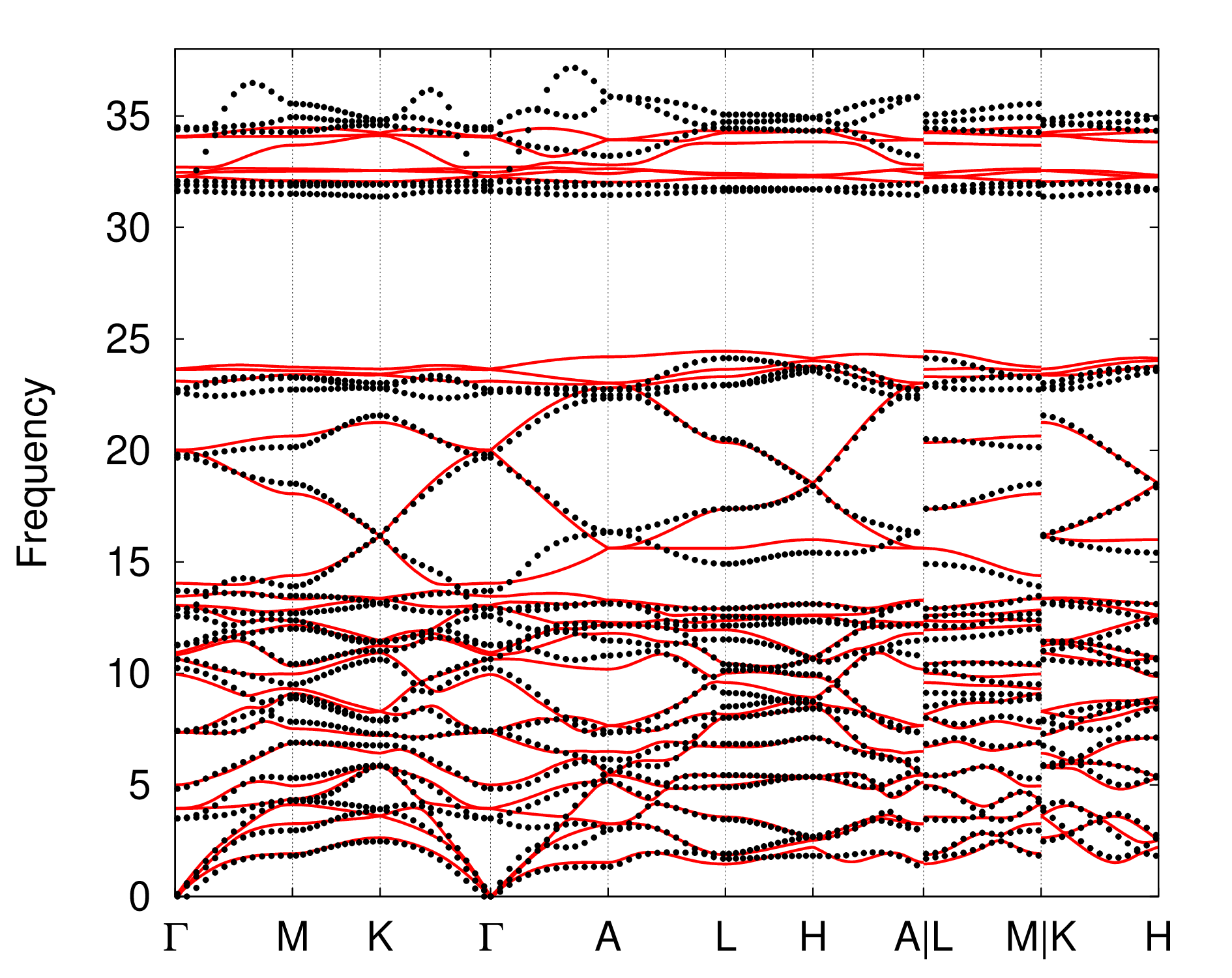}}
\subfigure[]{\includegraphics[width=1.7in, height=1.7in, keepaspectratio=false]{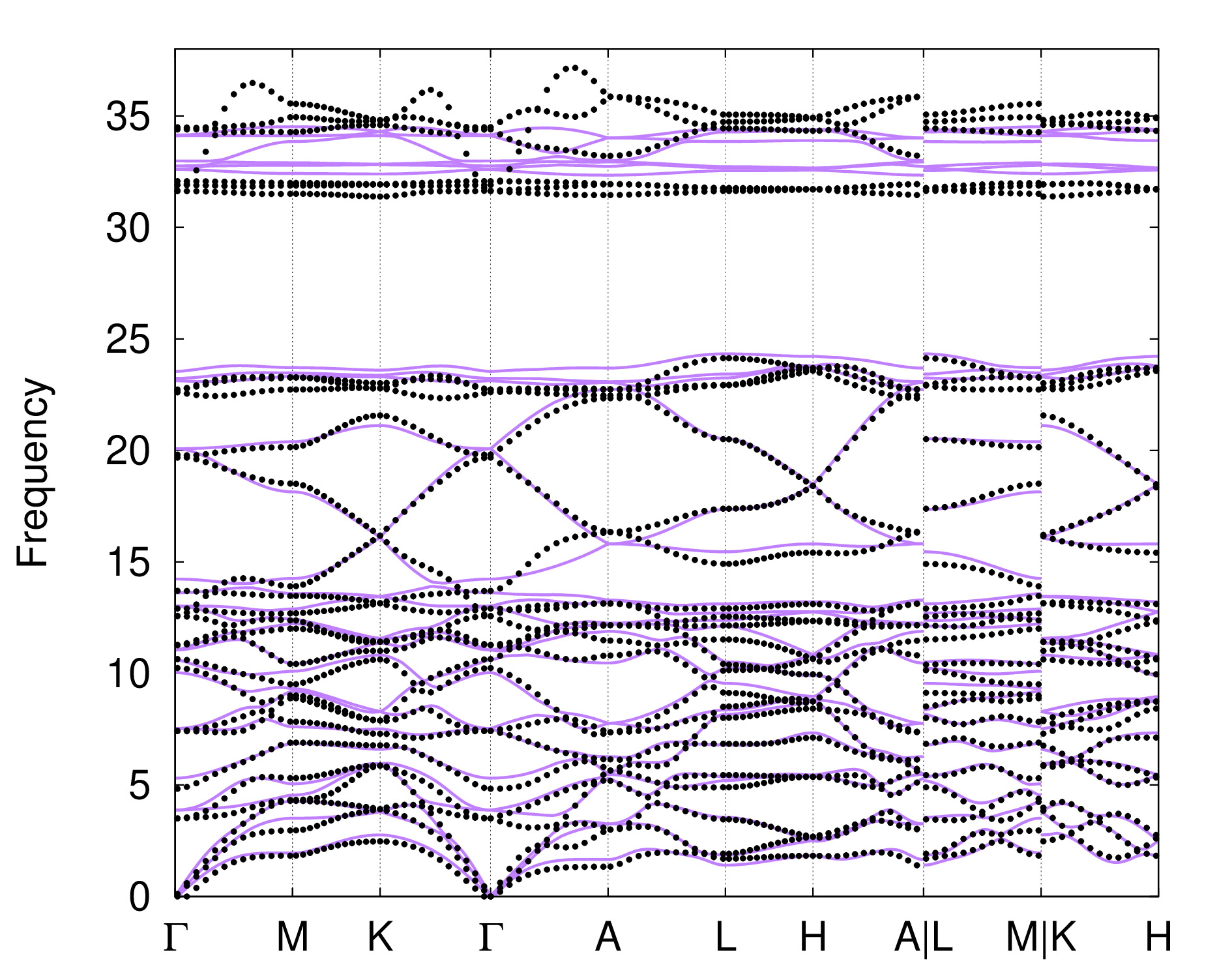}}
\subfigure[]{\includegraphics[width=1.7in, height=1.7in, keepaspectratio=false]{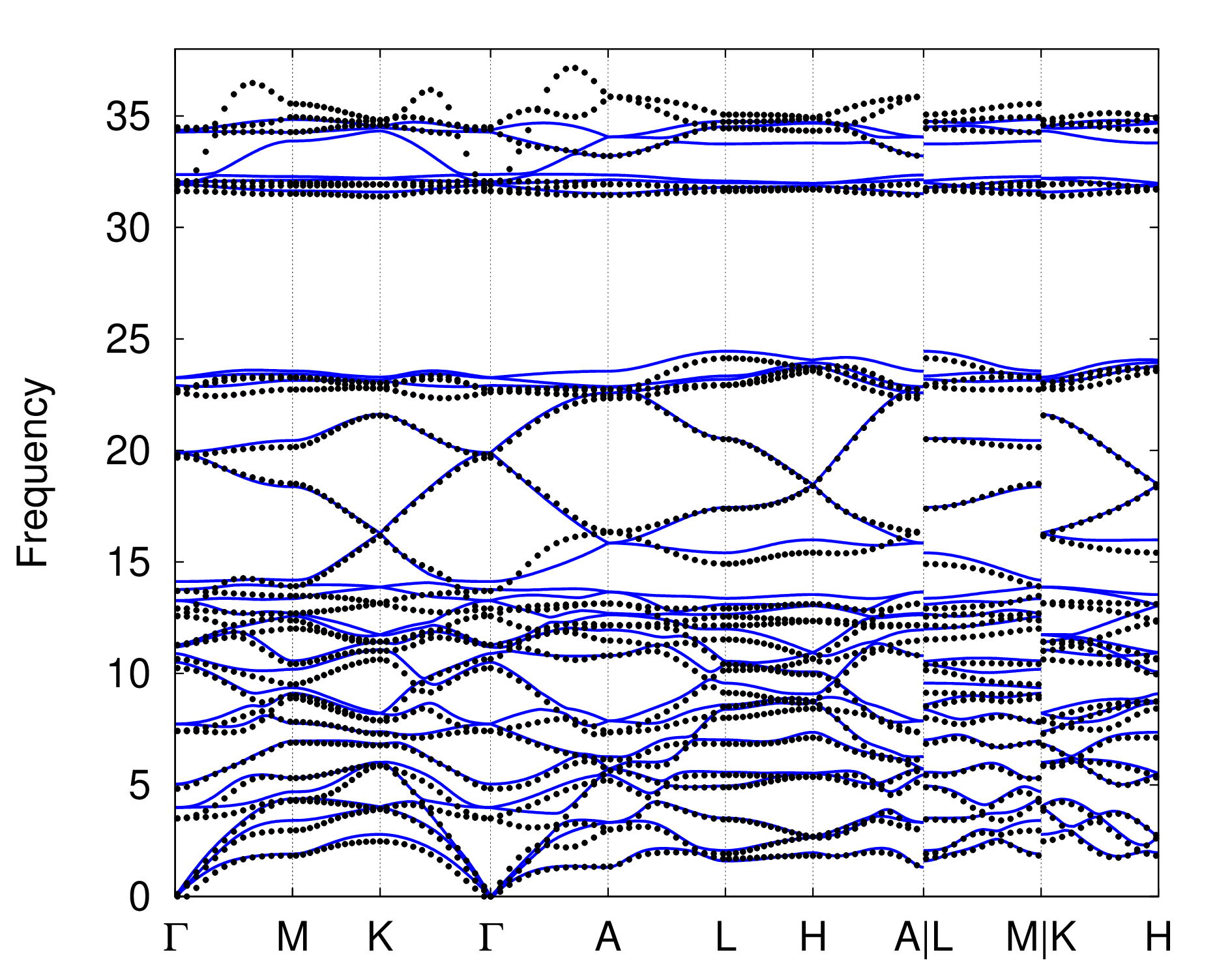}}
\subfigure[]{\includegraphics[width=1.7in, height=1.7in, keepaspectratio=false]{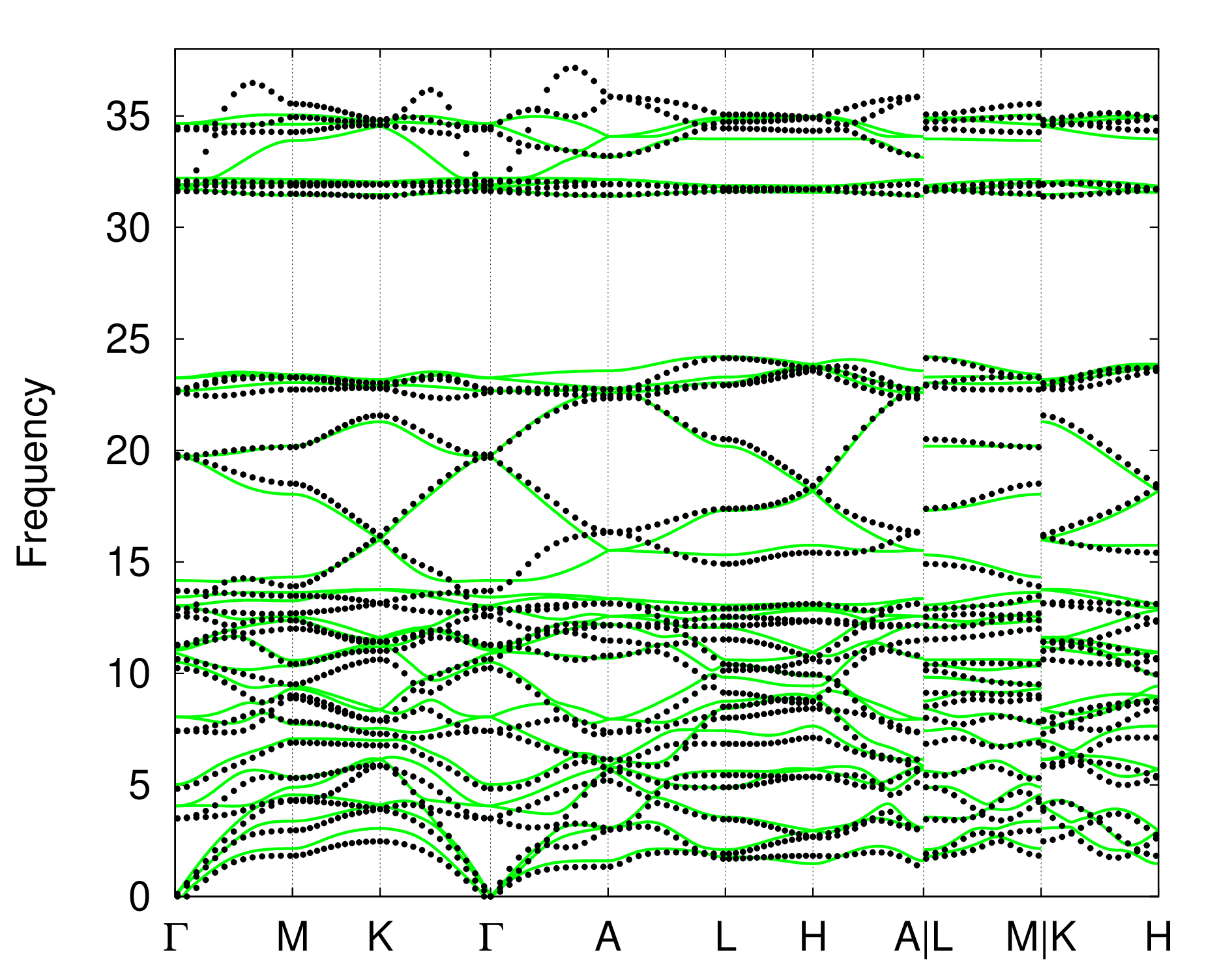}}
\subfigure[]{\includegraphics[width=1.7in, height=1.7in, keepaspectratio=false]{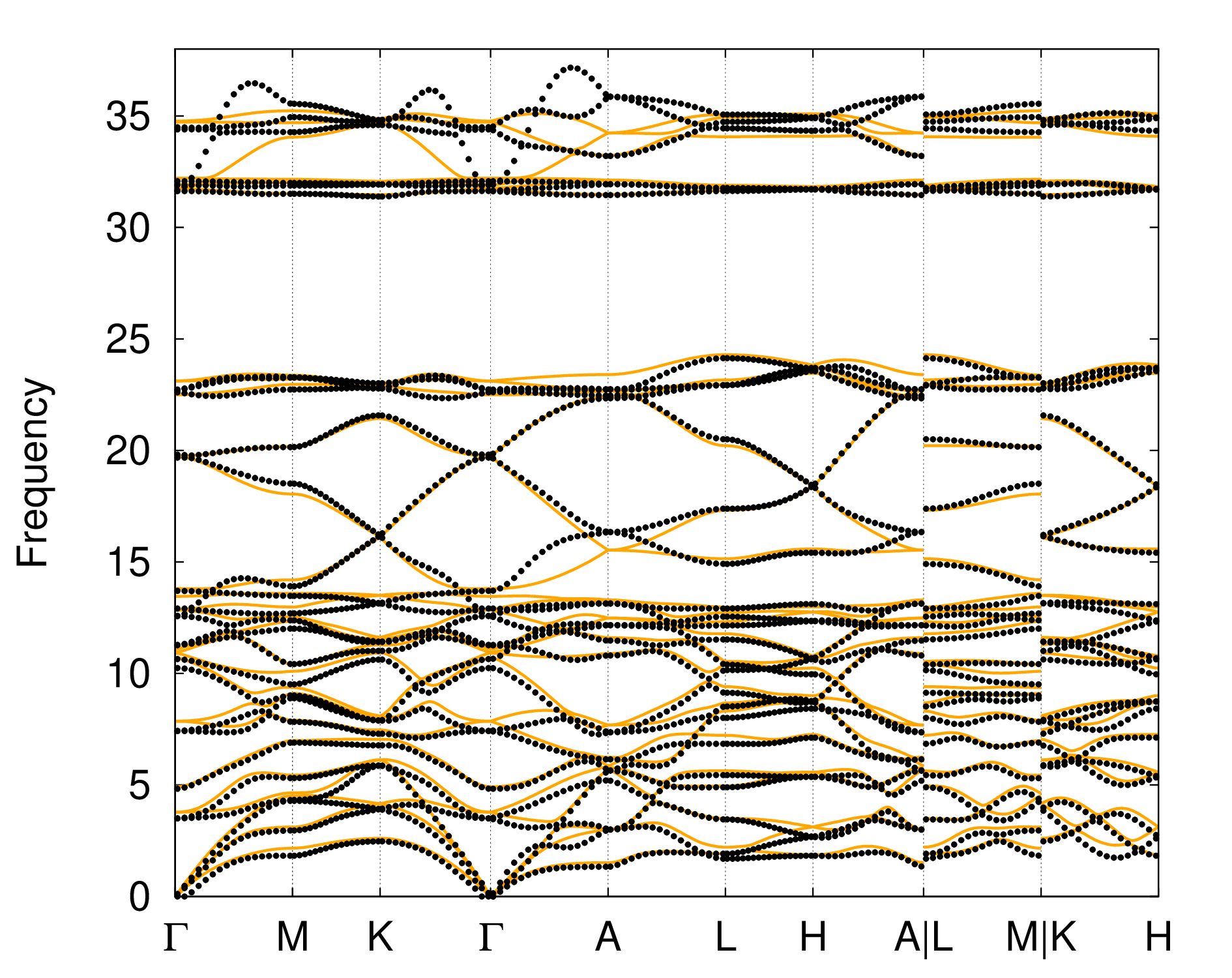}}
\caption{\label{fig:PhononsUncertainty} Phonon spectra of SiO$_2$ $\alpha$-quartz predicted by MTP$_1$, MTP$_1$+QEq, MTP$_2$, MTP$_2$+QEq and MTP$_3$ (the figures (a), (b), (c), (d), and (e), respectively) compared to the reference DFT phonon spectrum (black circles). Each ensemble of potentials, generally, shows a good agreement with the reference DFT data except for the very high frequencies. MTP$_1$ and MTP$_1$+QEq appear to be least accurate among all the ensembles of potentials, adding QEq to MTP does not improve the spectrum.}
\end{center} \end{figure}

Next we test how well the potentials reproduce the phonon spectrum. Again, we compare average phonon spectra to the reference DFT spectrum, i.e., we take the averaged phonon spectrum for each ensemble of potentials rather than one spectrum from one potential.
We used the PHONOPY open-source package \cite{togo2015-PHONOPY} to plot the spectra. The results shown in Figure \ref{fig:PhononsUncertainty} were obtained with the $4 \times 4 \times 4$ supercell of the nine-atom primitive unit cell. The k-path for this system was $\Gamma$-M-K-$\Gamma$-A-L-H-A$|$L-M$|$K-H (see, e.g., \cite{jain2013-MaterialsProject}), where $\Gamma$, M, K, A, L and H are the high-symmetry points in the Brillouin zone.
Each ensemble of potentials, generally, shows a good agreement with the reference DFT data except for the very high frequencies, but, as it was expected, MTP$_1$ and MTP$_1$+QEq appear to be the least accurate among all the ensembles of potentials, and adding QEq to MTP does not improve the spectrum.
%The phonon spectrum obtained on the MTP shows a very good agreement with the DFT phonon spectrum. The phonon spectra obtained on the SW potential reproduced the DFT phonon spectrum only qualitatively, especially for the high frequencies. The SW+QEq model demonstrated the worst reproduction of the reference spectrum. 
%Thus, these models describe vibrational properties of $\alpha$-quartz SiO$_2$ worse than MTP.

\begin{figure}[h!] \begin{center}
\subfigure[]{\includegraphics[width=2.6in, height=2.4in, keepaspectratio=false]{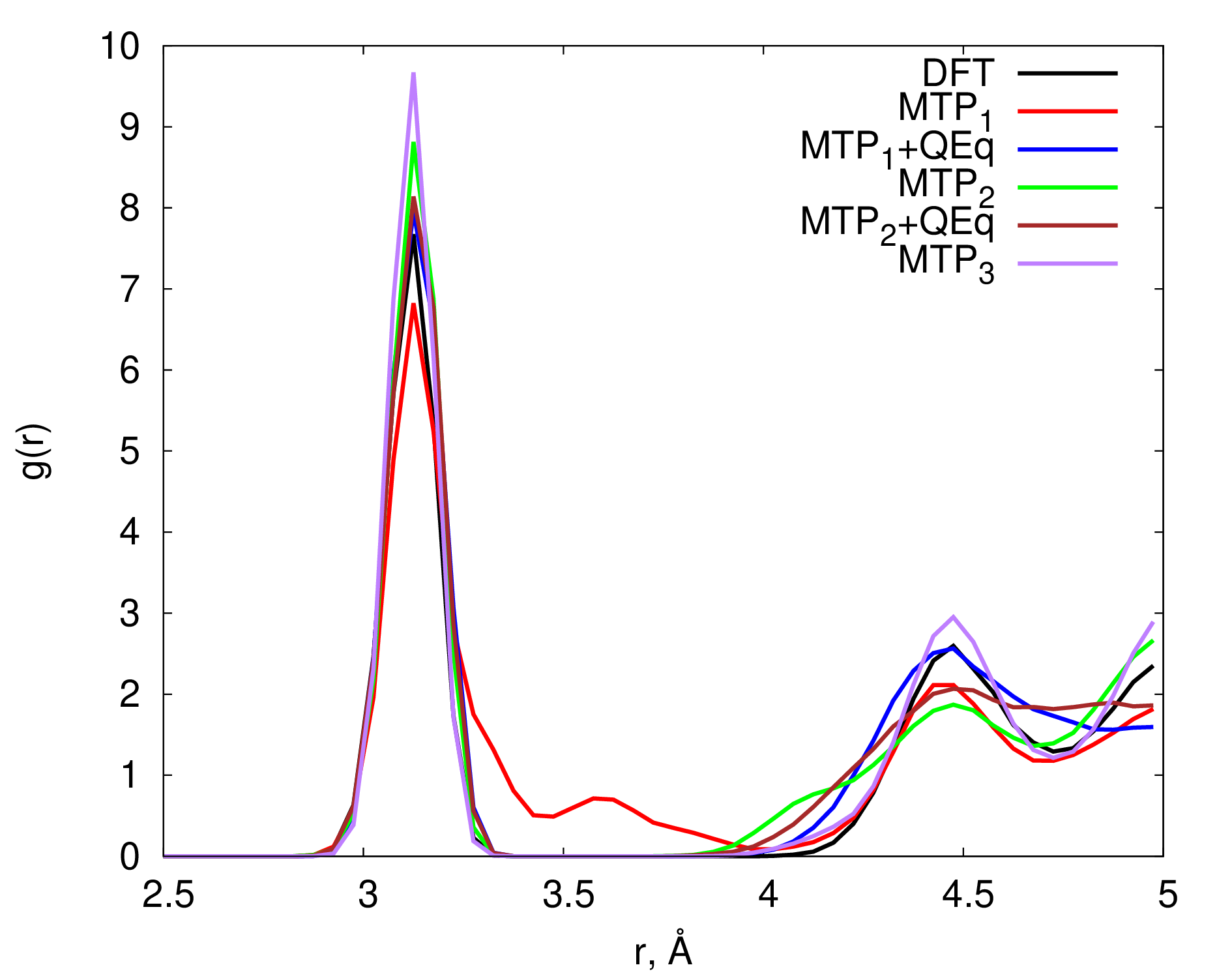}}
\subfigure[]{\includegraphics[width=2.6in, height=2.4in, keepaspectratio=false]{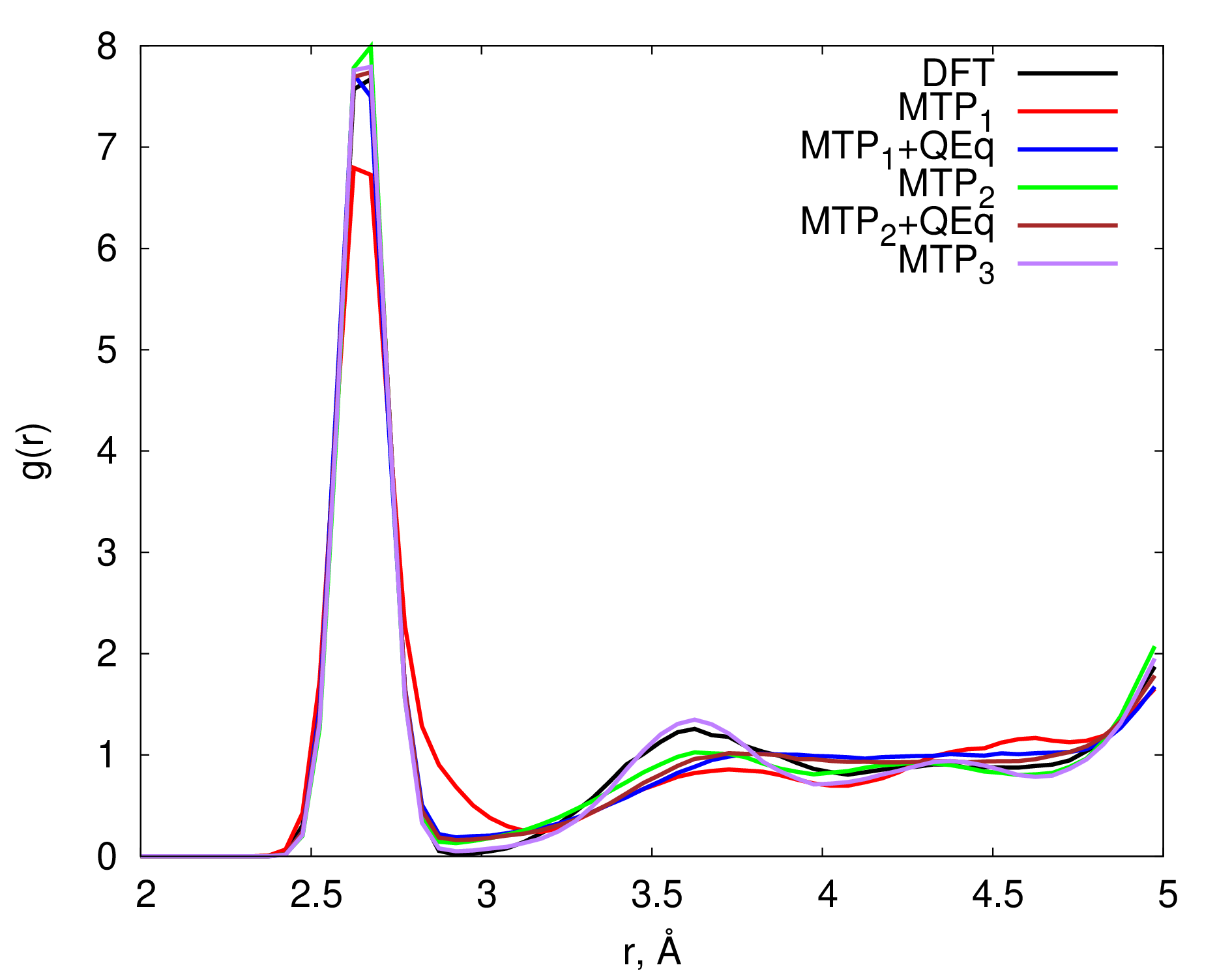}}
\caption{\label{fig:RDFs_Uncertainty} Radial distribution function (a) for Si--Si, and (b) O--O, computed at $T=300$ K. MTP$_3$ is in the best agreement with DFT, MTP$_1$ has the worst correspondence to the reference RDFs, the results of the rest three models are close to each other.}
\end{center} \end{figure}

Finally, we compare the radial distribution functions.
The RDF calculations were performed in a $2 \times 2 \times 2$ supercell with 72 atoms.
To obtain the RDFs for the five ensembles of potentials we ran molecular dynamics on LAMMPS \cite{plimpton1995-LAMMPS} sampling an NVT ensemble with $T = 300\,$K and time step of 1 fs.
The reference RDFs were obtained by running molecular dynamics on VASP.
The RDFs are plotted in Figure \ref{fig:RDFs_Uncertainty}.
MTP$_3$ showed the best correspondence with the reference RDFs, this is the only potential which correctly described all the peaks of O-O RDF. RDFs predicted with MTP$_1$ have the worst agreement with the reference RDFs among all the ensembles of the potentials, the rest three potentials have demonstrated the close accuracy to each other in description of the DFT RDFs.
%All the three models correctly predict the Si--Si RDF, with SW+QEq being slightly more accurate than the other two.
%The MTP reproduced the peaks just qualitatively, but not quantitatively whereas the SW potentials predicted the second peak (near $r = 4.5 \angstrom$) incorrectly.
%On the contrary, for the O--O RDFs, the MTP is in a good agreement with DFT, while SW and SW+QEq did not describe the second peak (near $r = 3.5 \angstrom$) correctly.

\section{Conclusion} \label{Discussion}

In this work we investigated two strategies for improving the accuracy of a machine-learning interatomic potential, namely adding more fitting parameters to it and adding a charge-equilibration model to it.
To that end we tested the MTP potentials \cite{shapeev2016-mtp,gubaev2018-chemoinformatics,gubaev2018-mtp-multicomponent} with increasing number of fitting parameters and MTP combined with the charge-equilibration model (MTP+QEq).
In order to make a meaningful comparison we assessed the uncertainty of predictions of each potential.
The uncertainty was due to the fact that typically the parameters of the fitted potentials are only near a local optimum, and an optimization routine typically finds some random local optimum.
Our conclusion is that adding more parameters to MTP does improve its predictive accuracy and reduces uncertainty of its predictions, whereas supplementing MTP with a charge-equilibration model does not reduce the error and often increases the uncertainty of the predictions.
We thus conclude that the QEq model could not, at least in a straightforward manner (i.e., by local optimization of model parameters), improve machine-learning potentials.
	
%Our main conclusion is that MTP has the lowest error and uncertainty among the three potentials and it is sufficient to use a local optimization algorithm for the parametrization of MTP.
%However, adding the QEq model to MTP did not make any improvement over the single MTP potential.
%In order to understand the underlying reason, we tested and compared the SW and SW+QEq models.
%We found that the uncertainties of their predictions were rather high and many instances of such potentials yield unphysical results (such as negative C$_{11}$ or C$_{33}$ elastic constants or negative vacancy formation energies).
%We emphasize that the empirical potentials are usually fitted directly to the quantities of interest (such as elastic constants, defect formation energies, etc.) with a global optimization algorithm.
%This indicates that a global optimization algorithms (or maybe some other, more sophisticated algorithm) have to be used to fit MTP+QEq in order to see improvement over MTP, while MTP itself shows a significant improvement over classical interatomic potentials.
%It should be added that a charge-equilibration model is typically more computationally expensive than even a machine-learning potential (see the Supplemental Materials for details).
%It is, thus, not trivial to practically improve machine-learning potentials by adding a charge equilibration model.

\section{Acknowledgements}

The authors thank our colleague, Konstantin Gubaev, for giving advance access to the code implementing MTP. I.N. also thanks Dmitry Aksenov for valuable advices concerning the usage of DFT (VASP) and application of PHONOPY to the computations of phonon spectra. The work was supported by the Russian Science Foundation (grant number 18-13-00479).  

\section{Data availability}

The data
%required to reproduce our results
used for the fitting of the interatomic potentials
are available to download from \url{http://gitlab.skoltech.ru/Novikov/SiO2_training_set}.

\bibliographystyle{elsarticle-num}
\bibliography{article}

\end{document}